\documentclass[aps,pra,amssymb,amsmath,eqsecnum,nofootinbib]{revtex4}
\usepackage{graphicx}

\newtheorem{definition}{Definition}[section]
\newtheorem{theorem}{Theorem}[section]
\newtheorem{proposition}{Proposition}[section]

\newtheorem{corollary}{Corollary}[section]
\newtheorem{conjecture}{Conjecture}[section]
\newtheorem{remark}{Remark}[section]
\newtheorem{axiom}{AXIOM}[section]

\newenvironment{hypothesis}{HP: \begin{center}} {\end{center}}
\newenvironment{thesis}{TH: \begin{center}} {\end{center}}
\newenvironment{proof}{\begin{center}PROOF: \end{center}} {$ \blacksquare $}
\newtheorem{example}{Example}[section]
\begin{document}
\title{Von Neumann Uniqueness Theorem doesn't hold in Hyperbolic Quantum
Mechanics}
\author{Andrei Khrennikov, Gavriel Segre}
\email{Andrei.Khrennikov@msi.vxu.se, Gavriel.Segre@msi.vxu.se}
\affiliation{International Center for Mathematical Modelling in
Physics and Cognitive Sciences, University of V\"{a}xj\"{o},
S-35195, Sweden}
\begin{abstract}
It is shown that Von Neumann Uniqueness Theorem doesn't hold in
Hyperbolic Quantum Mechanics
\end{abstract}

\maketitle
\newpage
\tableofcontents
\newpage
\section{Introduction}
Following  Adler's \cite{Adler-95}  formalization of Feynman's
basic observations \cite{Feynman-Hibbs-65} concerning quantum
probabilities, let us recall that a key feature of quantum
probabilities consists in that they don't obey the usual
\emph{formula of probabilities' composition}:
\begin{equation} \label{eq:classical probabilities' composition}
    P_{c \, a} \; = \; \sum_{b}  P_{c \, b} \cdot P_{b \, a}
\end{equation}
but a \emph{formula for probabilities amplitudes's composition}:
\begin{equation} \label{eq:quantum probabilities' composition}
    \Phi_{c \, a } \; = \; \sum_{b} \Phi_{c \, b } \cdot \Phi_{b \, a }
\end{equation}
where the probabilities amplitudes $\Phi$'s take value on a finite
dimensional algebra A over $ {\mathbb{R}} $ \cite{Shafarevich-97}
on which a \emph{modulus function} $ N : A \mapsto {\mathbb{R}}$
is defined such that:
\begin{eqnarray}
  P_{b \, a} \; &=& \;   N^{2} ( \Phi_{b \, a })  \\
  P_{c \, b} \; &=& \;   N^{2} ( \Phi_{c \, b })  \\
  P_{c \, a} \; &=& \;   N ^{2}( \Phi_{c \, a })
\end{eqnarray}
where both the algebra A and the modulus function has to be
determined imposing reasonable physical and mathematical
constraints.

From a mathematical side it is natural to require that N is a norm
over A.

From a physical side the imposition of the Correspondence
Principle requires that, in the absence of quantum interference
effects, probability amplitude superposition (i.e. eq.
\ref{eq:quantum probabilities' composition}) should reduce to
probability superposition (i.e. eq. \ref{eq:classical
probabilities' composition}). This leads (cfr. \cite{Adler-95} for
details) to the condition that the norm N has to be
multiplicative.

One has that:
\begin{theorem} \label{th:Albert's Theorem}
\end{theorem}
ALBERT'S THEOREM

\begin{hypothesis}
\end{hypothesis}
\begin{center}
    A  finite dimensional algebra with unit over $ {\mathbb{R}} $
\end{center}
\begin{center}
  N multiplicative norm over A
\end{center}

\begin{thesis}
\end{thesis}
\begin{equation*}
    A \; \in \; \{ {\mathbb{R}} , {\mathbb{C}} , {\mathbb{H}} , {\mathbb{O}}  \}
\end{equation*}
where $ {\mathbb{H}} $ is the (noncommutative) algebra of
Hamilton's \emph{quaternions} and $ {\mathbb{O}} $ is the
(noncommutative and nonassociative) algebra of Cayley's
\emph{octonions} whose definition we briefly recall.

The generic element of an (n+1)-dimensional algebra with unit A
may be expressed as:
\begin{equation}
    \Phi \; = \; \sum_{i=0}^{n} r_{i} e_{i} \; \;
\end{equation}
where $ e_{0}=1 , \cdots , e_{n} $  are the basis elements  of the
algebra obeying multiplication law:
\begin{equation}
    e_{i} \cdot  e_{j} \; = \; \sum_{k=0}^{n} f_{ijk} e_{k} \; \;
    i,j = 0 , \cdots , n
\end{equation}
with the  real-valued structure constant $ f_{ijk} $'s obeying the
following constraints:
\begin{equation}
    f_{0ij} \; = \; \delta_{ij} \; \;  i,j = 0 , \cdots , n
\end{equation}
\begin{equation}
    f_{i0j} \; = \; \delta_{ij} \; \;  i,j = 0 , \cdots , n
\end{equation}
that may be immediately derived imposing that:
\begin{equation}
    e_{i} \cdot e_{0} \; = \; e_{0} \cdot e_{i} \; = \; e_{0} \;
    \; i = 0 , \cdots , n
\end{equation}
The algebra of quaternions ${\mathbb{H}}$ corresponds to the case
$n=3 $ and:
\begin{equation}
    e_{i} \cdot  e_{j}  \; = \; - \delta_{ij} + \sum_{k=1}^{3} \epsilon_{ijk}
    e_{k}
\end{equation}
where  $ \epsilon_{ijk}  $ is the Levi Civita's tensor, i.e. the
totally antisymmetric tensor with $ \epsilon_{123} \, = \, 1$.

A multiplicative norm on ${\mathbb{H}}$  is given by:
\begin{equation}
    N( \sum_{i=0}^{3} r_{i} e_{i}  ) \; := \; \sqrt{ \sum_{i=0}^{3} r_{i}^{2} }
\end{equation}
The algebra of octonions ${\mathbb{O}}$ corresponds to the case
$n=7 $ and:
\begin{equation}
    e_{i} \cdot  e_{j}  \; = \; - \delta_{ij} + \sum_{k=1}^{3} \gamma_{ijk}
    e_{k}
\end{equation}
where $ \gamma_{ijk} $ is the totally antisymmetric tensor such
that:
\begin{equation}
  \gamma_{123} \; = \;  \gamma_{246} \; = \; \gamma_{435} \; = \;
  \gamma_{367} \; = \; \gamma_{651} \; = \;  \gamma_{572} \; = \;
  \gamma_{714} \; = \; 1
\end{equation}
A multiplicative norm on ${\mathbb{O}}$ is given by:
\begin{equation}
    N( \sum_{i=0}^{7} r_{i} e_{i}  ) \; := \; \sqrt{ \sum_{i=0}^{7} r_{i}^{2} }
\end{equation}

\medskip

 Let us observe, anyway, that from a physical viewpoint
it seems rather natural to require that:
\begin{equation}
    \Phi_{a \, b} \cdot  \Phi_{b \, c} \; = \; \Phi_{b \, c} \cdot
     \Phi_{a \, b}
\end{equation}
that leads to the constraint that A has to be commutative; so, by
Albert Theorem,  $ A \in \{ {\mathbb{R}} , {\mathbb{C}} \} $.

Since $ {\mathbb{R}} $  is to strict to allow all the
superpositions of states observed experimentally, it follows that
Nature has chosen $ A = {\mathbb{C}} $.

Let us now observe that, with the exception of the non-associative
octonions' algebra $  {\mathbb{O}} $, the algebras allowed by
Albert's Theorem are finite-dimensional real Clifford algebras
\cite{Gurlebeck-Sprossig-97}:
\begin{eqnarray}
  {\mathbb{R}} \; &=& \; Cl_{0,0} \\
  {\mathbb{C}} \; &=& \; Cl_{0,1} \\
  {\mathbb{H}} \: &=& \; Cl_{0,2} \\
\end{eqnarray}
To understand better the structural properties of the choice of
$Cl_{0,1} $ made by Nature, it may be interesting to investigate
how the mathematical structure of Quantum Mechanics is modified by
the ansatz:
\begin{equation}
    Cl_{0,1} \; \mapsto \; Cl_{1,0}
\end{equation}
corresponding, in terms of the structure constants, to the ansatz:
\begin{equation}
    f_{111} = 0 \, f_{110} = 1 \; \mapsto \;  f_{111} = 0 \, f_{110} = -1
\end{equation}
 Since, as we have shown in \cite{Khrennikov-Segre-05a}, there
are many reasons to call $ {\mathbb{G}} \; := \; Cl_{1,0} $ the
\emph{hyperbolic algebra}, we  will denote the $Cl_{1,0}$-Quantum
Mechanics as Hyperbolic Quantum Mechanics.

Such a mathematical theory emerged in the research of one of the
authors \cite{Khrennikov-03a}, \cite{Khrennikov-03b},
\cite{Khrennikov-04} and of other scientists (see \cite{Kocik-99},
\cite{Xuegang-00}, \cite{Zheng-Xuegang-04},
\cite{Rochon-Tremblay-04}, \cite{Ulrych05a}, \cite{Ulrych05b},
\cite{Ulrych05c} and references therein).

Let us observe, first of all, that the modulus function:
\begin{equation}
    N( \sum_{i=0}^{1} r_{i} e_{i}  ) \; := \; \sqrt{ \sum_{i=0}^{1} r_{i}^{2} }
\end{equation}
is a norm, though not multiplicative.

\begin{remark}
\end{remark}
 We would like to stress, from the physical point of view, that we
are in no way claiming that Quantum Mechanics, as a physical
theory, is wrong or has to be modified.

We are simply analyzing an alternative mathematical theory whose
structure could allow to get some insight of the $
Cl_{0,1}$-choice made by Nature
\newpage
\section{The hyperbolic algebra}
Let us define the hyperbolic algebra as the ring $ {\mathbb{G}} $
of numbers of the form $ x + j y $, where $ x,y \in {\mathbb{R}} $
while j, called the hyperbolic imaginary unit, is such that $
j^{2} = +1 $.

The elements of such an algebra has been called in the
mathematical literature with different names (cfr.
\cite{Jancevic-96} and references therein): \emph{hyperbolic
numbers}, \emph{double numbers}, \emph{split complex numbers},
\emph{perplex numbers}, and \emph{duplex numbers}.

We will call them \emph{hyperbolic numbers}  and we will refer to
j as to the \emph{hyperbolic imaginary unit}.

The complex field ${\mathbb{C}}$ and the hyperbolic ring
${\mathbb{G}}$ are the two bidimensional Clifford algebras
\cite{Khrennikov-Segre-05a}:
\begin{eqnarray}
  {\mathbb{C}} \; &=& \; Cl_{0,1} \\
  {\mathbb{G}} \; &=& \; Cl_{1,0}
\end{eqnarray}

Given $ z = x + j y  \in  {\mathbb{G}}  $:
\begin{definition}
\end{definition}
CONJUGATE OF z:
\begin{equation*}
    \bar{z} \; := \; x - j y
\end{equation*}
\begin{definition}
\end{definition}
NORM OF z:
\begin{equation*}
     \| z \| := \sqrt{ x^{2} + y^{2} }
\end{equation*}
\begin{definition}
\end{definition}
LIGHT CONE OF $ z $:
\begin{equation*}
    V_{z} \; := \; \{ z' = x' + j y' \in {\mathbb{G}} \, : \, y' =
    \pm (x'-x) + y \}
\end{equation*}
One has that:
\begin{theorem}
\end{theorem}
\begin{itemize}
    \item $ {\mathbb{G}} $ is a commutative ring
    \item $ {\mathbb{G}} $ is not a field
\end{itemize}
\begin{proof}
\begin{enumerate}
    \item by definition the addition and the multiplication in $
    {\mathbb{G}}$ are commutative and associative, the
    multiplication is distributive with respect to the addition, there exists a null element 0 with respect to addition, there
    exists an identity element 1 with respect to multiplication
    and every element of $
    {\mathbb{G}}$ has an additive inverse
    \item
    given $ z = x + j y \in {\mathbb{G}} $ one has  formally that:
\begin{equation}
    \frac{1}{z} \; = \; \frac{1}{x + j y} = \frac{x}{x^{2}-y^{2}}
    - j \frac{y}{x^{2}-y^{2}}
\end{equation}
and hence:
\begin{equation}
  \exists z^{-1} \; \Leftrightarrow \; z \notin V_{0}
\end{equation}
So not every nonzero element of  ${\mathbb{G}} $ has a
multiplicative inverse and hence  ${\mathbb{G}} $ is not a field
\end{enumerate}
\end{proof}

One has clearly that:
\begin{proposition}
\end{proposition}
\begin{equation*}
    \| z \|^{2} \; \neq \; z \bar{z} \; \; \forall z \in
    {\mathbb{G}} - {\mathbb{R}}
\end{equation*}
\begin{proof}
 Given  $ z = x + j y \in
    {\mathbb{G}} - {\mathbb{R}}  $ one has that
 \begin{equation}
   \| z \|^{2} \; = \;  x^{2} + y^{2} \; \neq \; z \bar{z} \; = \;  x^{2} - y^{2}
\end{equation}
\end{proof}
\newpage
\section{Hyperbolic Hilbert Spaces}
Let us start introducing the following:
\begin{definition}
\end{definition}
HYPERBOLIC LINEAR SPACE:

a triple $ ( V , + , \cdot ) $ where V is a set while $ + : V
\times V \mapsto V $ and $ \cdot : {\mathbb{G}} \times V \mapsto V
$ are such that:
\begin{equation*}
    u + v = v + u \; \; \forall u,v \in V
\end{equation*}
\begin{equation*}
    (u+v)+w \; = \; u + ( v+w) \; \; \forall u,v,w \in V
\end{equation*}
\begin{equation*}
    \exists 0 \in V \: : \; u+0 = u \; \; \forall u \in V
\end{equation*}
\begin{equation*}
    \lambda ( u + v ) \; = \lambda u + \lambda v \; \; \forall u,v \in
    V, \forall \lambda \in {\mathbb{G}}
\end{equation*}
\begin{equation*}
    ( \lambda + \mu ) u \; = \; \lambda u + \mu u \; \; \forall u, \in
    V, \forall \lambda , \mu  \in {\mathbb{G}}
\end{equation*}
\begin{equation*}
    1 u \; = \; u \; \; \forall u \in V
\end{equation*}
We can then introduce the following:
\begin{definition}
\end{definition}
HYPERBOLIC INNER PRODUCT SPACE:

a quatruple $ ( V ,+ , \cdot , ( \cdot , \cdot) )$  such that:
\begin{itemize}
    \item $ ( V ,+ , \cdot ) $ is an hyperbolic linear space
    \item $ ( \cdot , \cdot) : V \times V \mapsto {\mathbb{G}} $ is such
    that:
\begin{equation*}
    ( u , v+w) \; = \; ( u , v) + (u , w) \; \; \forall u,v,w \in V
\end{equation*}
\begin{equation*}
    ( u , \lambda v ) \; = \; \lambda ( u,v) \; \; \forall u,v \in
    V, \forall \lambda   \in {\mathbb{G}}
\end{equation*}
\begin{equation*}
    (u,v) \; = \; \overline{(v,u)}  \; \; \forall u,v \in
    V
\end{equation*}
\end{itemize}
\begin{example}
\end{example}
Let $ {\mathbb{G}}^{n} $ denote the set of all n-ples of
hyperbolic numbers; given $ x = ( x_{1}, \cdots , x_{n} ) $, $y =
( y_{1} , \cdots , y_{n} ) \in   {\mathbb{G}}^{n} $ define:
\begin{equation}
    (x , y) \; := \; \sum_{i=1}^{n} \bar{x}_{i} y_{i}
\end{equation}
$ ({\mathbb{G}}^{n} , ( \cdot , \cdot )) $ is then an hyperbolic
inner product space

\smallskip

Given an hyperbolic inner-product space $  ( V ,+ , \cdot , (
\cdot , \cdot) )$ and a linear operator $ U : V \mapsto V $:
\begin{definition}
\end{definition}
U IS UNITARY:
\begin{equation*}
    ( U x , U y) \; = \; ( x , y ) \; \; \forall x,y \in V
\end{equation*}

\begin{definition}
\end{definition}
HYPERBOLIC NORMED LINEAR SPACE:

a  couple $ ( V , \| \cdot \| ) $ where:
\begin{itemize}
    \item V is an hyperbolic linear space
    \item $ \| \cdot \| $ is a map $ \| \cdot \| : {\mathbb{G}} \mapsto {\mathbb{R}}
    $ such that:
\end{itemize}
\begin{equation*}
  \| v \| \; \geq \; 0 \; \; \forall v \in V
\end{equation*}
\begin{equation*}
   \| v \| \; = \; 0  \; \; \Leftrightarrow \; \; v = 0
\end{equation*}
\begin{equation*}
 \exists c \in {\mathbb{R}}_{+} \; : \; ( \| \alpha  v \| \; \leq  \; c  \| \alpha \| \| v \| \; \; \forall v \in
  V , \forall \alpha \in {\mathbb{G}} )
\end{equation*}
\begin{equation*}
    \| u + v \| \; \leq \; \| u \| + \| v \| \; \; \forall u , v \in
     V
\end{equation*}
Given two hyperbolic normed linear spaces $ ( V_{1} , \| \cdot
\|_{1} ) $ and $ ( V_{2} , \| \cdot \|_{2} ) $ and a linear
operator $ T : V_{1} \mapsto V_{2} $ :
\begin{definition}
\end{definition}
T IS BOUNDED:
\begin{equation*}
    \| T \| \; := \sup_{ \| v \|_{1} = 1 } \| T v \|_{2} \in {\mathbb{R}}
\end{equation*}

\begin{definition}
\end{definition}
HYPERBOLIC BANACH SPACE

an hyperbolic normed linear space  $( V , \| \cdot \| ) $ which is
complete as a metric space in the induced metric $ d( u , v ) :=
\| u - v \| $.

\smallskip

\begin{definition}
\end{definition}
HYPERBOLIC HILBERT SPACE

a triple $ ( V , ( \cdot , \cdot ) , \| \cdot \| ) $ such that:
\begin{itemize}
    \item $ ( V , ( \cdot , \cdot )) $ is an hyperbolic
    inner-product space
    \item $ ( V , \| \cdot \| ) $ is an hyperbolic Banach space
    \item
\begin{equation*}
   \exists c \in {\mathbb{R}}_{+} \; : \; ( \| ( u , v ) \| \; \leq \; c \| u \| \| v \| \; \; \forall u,v \in
    V )
\end{equation*}
\end{itemize}

\smallskip

\begin{remark}
\end{remark}
Let us observe that all the introduced notions of the form
"hyperbolic x", with x = linear space, normed linear space, Banach
space, Hilbert space, has not to be intended as particular cases
of the respective notion x:

since $ { \mathbb{G}} $ is not a field, an hyperbolic linear space
is not a linear space but only a module over the ring $ {
\mathbb{G}} $ and so on.

\smallskip

\begin{example}
\end{example}
Given the hyperbolic inner product space $ ({\mathbb{G}}^{n} , (
\cdot , \cdot ))$ let us introduce the hyperbolic Banach space $ (
{\mathbb{G}}^{n} , \| \cdot \|) $ defined as:
\begin{equation}
    \| ( x_{1} + j y_{1} , \cdots ,  x_{n} + j y_{n}) \| \; := \; \sqrt{\sum_{i=1}^{n} (x_{i}^{2} + y_{i}^{2}) }
\end{equation}
Given $ x = (a_{1}+j b_{1} , \cdots , a_{n}+j b_{n} ) \in
{\mathbb{G}}^{n} $ and $ \alpha = c + j d   \in {\mathbb{G}} $ one
has that:
\begin{equation}
    \| \alpha x \| \; = \; \sqrt{ \sum_{i=1}^{n}( c^{2} a_{i}^{2} + c^{2} b_{i}^{2} + d^{2} a_{i}^{2} + d^{2} b_{i}^{2} + 4 a_{i} b_{i} c d  )}
\end{equation}
\begin{equation}
 \| \alpha\| \| x \| \; = \;  \sqrt{ \sum_{i=1}^{n}( c^{2} a_{i}^{2} + c^{2} b_{i}^{2} + d^{2} a_{i}^{2} + d^{2} b_{i}^{2}   )}
\end{equation}
Since:
\begin{equation}
    2 a_{i} b_{i} c d \; \leq \; c^{2} a_{i}^{2} + d^{2} b_{i}^{2}
\end{equation}
\begin{equation}
       2 a_{i} b_{i} c d \; \leq \; c^{2} b_{i}^{2} + d^{2} a_{i}^{2}
\end{equation}
it follows that:
\begin{equation}
    \| \alpha x \| \; \leq \; \sqrt{2} \| \alpha \| \| x \|
\end{equation}
Furthermore, given $ x = ( a_{1} + j b_{1} , \cdots, a_{n} + j
b_{n} ) , y  = ( c_{1}+j d_{1} , \cdots ,  c_{n}+j d_{n} ) \in
{\mathbb{G}}^{n} $ one has that:
\begin{equation}
    \| ( x , y ) \| \; = \; \sqrt{\sum_{i=1}^{n} (  a_{i}^{2} c_{i}^{2} + a_{i}^{2} d_{i}^{2} + b_{i}^{2} c_{i}^{2} + b_{i}^{2} d_{i}^{2} - 4 a_{i} b_{i} c_{i} d_{i}  )  }
\end{equation}
\begin{equation}
    \| x \| \| y \| \; = \; \sqrt{\sum_{i=1}^{n} (  a_{i}^{2} c_{i}^{2} + a_{i}^{2} d_{i}^{2} + b_{i}^{2} c_{i}^{2} + b_{i}^{2} d_{i}^{2}   )  }
\end{equation}
Since:
\begin{equation}
    - 2 a_{i} b_{i} c_{i} d_{i} \; \leq \;  a_{i}^{2} c_{i}^{2} + b_{i}^{2} d_{i}^{2}
\end{equation}
\begin{equation}
   - 2 a_{i} b_{i} c_{i} d_{i} \; \leq \; a_{i}^{2} d_{i}^{2} +  b_{i}^{2} c_{i}^{2}
\end{equation}
it follows that:
\begin{equation}
  \| ( x , y ) \| \; \leq \; \sqrt{2}  \| x \| \| y \|
\end{equation}

$ ({\mathbb{G}}^{n} , ( \cdot , \cdot ) , \| \cdot \|) $ is then
an hyperbolic Hilbert space.

\begin{example}
\end{example}
Define $ L^{2}(  {\mathbb{R}} , {\mathbb{G}} ) $ to be the set of
hyperbolic valued measurable functions on ${\mathbb{R}} $ that
satisfy
 $ \int_{-\infty}^{+ \infty} dx \| f(x) \|^{2} < + \infty $. Let us
 introduce:
 \begin{equation}
    (f , g) \; := \; \int_{-\infty}^{+ \infty} dx \bar{f}(x) g(x)
\end{equation}
and:
\begin{equation}
    \| \psi \| \; := \; \sqrt{\int_{- \infty}^{+ \infty} dx  \| \psi(x) \|^{2}}
\end{equation}
One has that:
\begin{equation}
    \| z \psi \| \; = \;  \sqrt{\int_{- \infty}^{+ \infty} dx  \|z \psi(x)
    \|^{2}} \; \leq \; \sqrt{2} \sqrt{\int_{- \infty}^{+ \infty} dx  \|z \|^{2} \| \psi(x)
    \|^{2}} \; = \; \sqrt{2} \| z \| \| \psi \| \; \; \forall z
    \in {\mathbb{G}} , \forall \psi \in L^{2}(  {\mathbb{R}} , {\mathbb{G}} )
\end{equation}
Furthermore one has that:
\begin{multline}
    \| ( f , g )  \| \; = \; \|  \int_{-\infty}^{+ \infty} dx   \bar{f}(x)
    g(x) \| \; \leq \; \int_{-\infty}^{+ \infty} dx \| \bar{f}(x)
    g(x) \| \;
    \leq \; \sqrt{2} \int_{-\infty}^{+ \infty} dx \| \bar{f}(x)
   \| \| g(x) \| \; \leq \\
   \leq \;  \sqrt{2} \sqrt{ \int_{-\infty}^{+ \infty} dx  \| f(x) \|^{2}}  \sqrt{ \int_{-\infty}^{+ \infty} dx  \| g(x) \|^{2}  }  \; = \; \sqrt{2} \| f \| \| g \|    \; \; \forall f,g \in L^{2}(  {\mathbb{R}} , {\mathbb{G}} )
\end{multline}
from which, using the fact that absolute convergence implies
convergence, it follows that:
\begin{equation}
  ( f , g ) \in {\mathbb{G}} \; \; \forall f,g \in L^{2}(  {\mathbb{R}} , {\mathbb{G}} )
\end{equation}
$ ( L^{2} ( {\mathbb{R}} , {\mathbb{G}}) , ( \cdot , \cdot ) , \|
\cdot \| ) $ is then an hyperbolic Hilbert space.

\smallskip

As to unbounded operators over an hyperbolic Hilbert space $
{\mathcal{H}} $ let us observe that, as in the analogous case of
unbounded operators over a  (complex) Hilbert space
\cite{Reed-Simon-80}, they will be usually defined only on a dense
linear subspace of $ {\mathcal{H}} $.

\newpage
 \section{Hyperbolic Quantum Mechanics}
Hyperbolic numbers emerged in the research of one of the authors
\cite{Khrennikov-04}, \cite{Khrennikov-03a}, \cite{
Khrennikov-03b} as the underlying number system of a mathematical
theory, "Hyperbolic Quantum Mechanics" formalized by the following
axioms:
\begin{axiom}\label{ax:states}
\end{axiom}

\emph{The pure states of an hyperbolic quantum systems are rays on
an hyperbolic Hilbert space $ {\mathcal{H}} $}

\medskip

\begin{axiom} \label{ax:observables}
\end{axiom}

\emph{Hyperbolic quantum mechanical observables are linear
operators on $  {\mathcal{H}} $ having real spectrum. The expected
value of the hyperbolic observable $ \hat{O} $ in a state $ \psi
\in {\mathcal{H}} $ such that $ ( \psi , \psi ) \neq 0 $ is given
by:}
\begin{equation}
    E_{\psi} (O) \; = \; \frac{(\psi , O \psi)  }{ (\psi , \psi ) }
\end{equation}

\begin{axiom} \label{ax:infinitesimal generators}
\end{axiom}

\emph{The evolution of a pure state $ \psi_{0} \in {\mathcal{H}} $
is described by the hyperbolic analogue of Schr\"{o}dinger's
equation:}
\begin{equation}
    j \frac{ d \psi(t) }{d t} \; = \; H \psi(t) \, , \, \psi(0) = \psi_{0}
\end{equation}

\newpage
\section{About Von Neumann Uniqueness Theorem}

Let us leave aside for a moment Hyperbolic Quantum Mechanics and
let us analyze the status of Von Neumann Uniqueness Theorem in
ordinary (complex) Quantum Mechanics.

Given an Hilbert space $ {\mathcal{H}} $, a dense linear subspace
$  {\mathcal{D}} $ of $ {\mathcal{H}} $ and two linear operators $
\hat{\tilde{Q}} , \hat{\tilde{P}} $ over $ {\mathcal{H}} $ we will
say that:
\begin{definition}
\end{definition}
$ \hat{\tilde{Q}} $ AND $  \hat{\tilde{P}} $ ARE A REPRESENTATION
OF THE CANONICAL COMMUTATION RELATION OVER $  {\mathcal{D}} $:
\begin{itemize}
    \item
\begin{equation*}
     {\mathcal{D}} \subseteq D( \hat{\tilde{Q}} ) \cap  D( \hat{\tilde{P}}
     ) \; , \; \hat{\tilde{Q}} {\mathcal{D}} \subseteq
     {\mathcal{D}} \; , \; \hat{\tilde{P}} {\mathcal{D}} \subseteq
     {\mathcal{D}}
\end{equation*}
\item
\begin{equation*}
     ([ \hat{\tilde{Q}}  ,  \hat{\tilde{P}} ])[ \psi ] \; := \;
    \hat{\tilde{Q}} [ \psi ] \hat{\tilde{P}}  [ \psi ]  -  \hat{\tilde{P}} [ \psi
    ] \hat{\tilde{Q}} [ \psi ]  \; = \; i \hat{I} [ \psi ] \; \;
    \forall \psi \in {\mathcal{D}}
\end{equation*}
\end{itemize}
where $ \hat{I} $ is the identity operator over  $ {\mathcal{H}}
$.

Introduced the following operators on $ L^{2}( {\mathbb{R}} ,
{\mathbb{C}}) $:
\begin{equation}
    ( \hat{Q}_{q} \psi ) ( q) \; := \; q \psi (q)
\end{equation}
\begin{equation}
     ( \hat{P}_{q} \psi ) ( q) \; := \; - i  \frac{d \psi (q)}{ d q}
\end{equation}
 (where i is the usual complex imaginary unit such that $ i^{2} =
- 1 $) defined as the closures of their restriction to the initial
domain $ {\mathcal{S}} ( {\mathbb{R}} , {\mathbb{C}}) $, it may be
easily verified that $ \hat{Q}_{q} $ and $ \hat{P}_{q} $ are a
representation of the \emph{Canonical Commutation Relation} called
the \emph{Schr\"{o}dinger representation}.

Von Neumann Uniqueness Theorem is often expressed in the Physics'
literature as the following:
\begin{conjecture} \label{con:naive version of Von Neumann Uniqueness Theorem}
\end{conjecture}
NAIVE VERSION OF VON NEUMANN UNIQUENESS THEOREM

\begin{hypothesis}
\end{hypothesis}
\begin{center}
  $ \hat{\tilde{Q}} $ and $  \hat{\tilde{P}} $ are a
  representation of the \emph{Canonical Commutation Relation} over the
  Hilbert space $ {\mathcal{H}} $
\end{center}
\begin{thesis}
\end{thesis}
\begin{equation*}
    \exists \hat{U} :  {\mathcal{H}} \mapsto  L^{2}( {\mathbb{R}} ,
{\mathbb{C}}) \text{ unitary } \; : \; \hat{\tilde{Q}} =
\hat{U}^{- 1} \hat{Q}_{q} \hat{U} \; and \;  \hat{\tilde{P}} =
\hat{U}^{- 1} \hat{P}_{q} \hat{U}
\end{equation*}
A mathematically more rigorous investigation allows anyway to
infer that \cite{Thirring-81}:

\begin{theorem} \label{th:naive version of Von Neumann Uniqueness
 Theorem is false}
\end{theorem}
\begin{center}
 Conjecture\ref{con:naive version of Von Neumann Uniqueness Theorem} is false
\end{center}
\begin{proof}
Let us consider the following Hilbert space $ ( l_{2} (
{\mathbb{C}} ) , ( \cdot , \cdot ) ) $:
\begin{equation}
  l_{2} ({\mathbb{C}} ) \; := \; \{ \{ x_{n} \}_{n=1}^{\infty} ,
  x_{n} \in {\mathbb{C}} \; \forall n \; : \; \sum_{n=1}^{\infty} |
  x_{n} |^{2} < \infty \}
\end{equation}
\begin{equation}
    ( \{ x_{n} \}_{n=1}^{\infty} ,  \{ y_{n} \}_{n=1}^{\infty} )
    \; := \; \sum_{n=1}^{\infty} \bar{x}_{n} y_{n}
\end{equation}
its dense linear subspace:
\begin{equation}
    {\mathcal{D}} \; := \; \{  \{ x_{n} \}_{n=1}^{\infty} \in   l_{2} (
{\mathbb{C}} ) \; : \; \sum_{n=1}^{\infty} x_{n} = 0 , \text{ with
only finitely many } x_{n} \neq 0 \}
\end{equation}
Given the infinite-dimensional matrices:
\begin{equation}
  \hat{\tilde{Q}} \; := \; diagonal ({\mathbb{N}})
\end{equation}
\begin{equation}
     \hat{\tilde{ P}} \; := -i \; \left(%
\begin{array}{ccccc}
  0 & -1 & - \frac{1}{2} & - \frac{1}{3} & \cdots \\
  1 & 0 &  -1 & - \frac{1}{2} & \cdots  \\
  \frac{1}{2} & 1 & 0 & -1 & \cdots \\
  \frac{1}{3} & \frac{1}{2} & 1 & 0 & \cdots \\
  \vdots & \vdots & \vdots & \vdots & \ddots \\
\end{array}%
\right)
\end{equation}
one has that $ \hat{\tilde{ Q}} $ and  $ \hat{\tilde{ P}} $ are a
representation over the dense linear subspace $ {\mathcal{D}} $ of
the \emph{Canonical Commutation Relation} unitarily inequivalent
to the Schr\"{o}dinger representation.

For other contra-examples see \cite{Summers-01} and references
therein
\end{proof}

Theorem\ref{th:naive version of Von Neumann Uniqueness
 Theorem is false} has led most of  the Mathematical Physics' community to
consider representations not of the\emph{ Canonical Commutation
Relation} but of the following \emph{Weyl relation}:
\begin{equation}
    \hat{V}_{1}(t) \hat{V}_{2}(s) \; = \; \exp ( i t s) \hat{V}_{2}(s)
    \hat{V}_{1}(t) \; \; \forall t , s \in {\mathbb{R}}
\end{equation}
of which the strongly continuous unitary groups $ \{ \exp ( i t
\hat{P}_{q} ) \}_{t \in {\mathbb{R}} } $ and $ \{  \exp ( i s
\hat{Q_{q}} ) \}_{s \in {\mathbb{R}} } $ are indeed a
representation, and to call Von Neumann Uniqueness Theorem  the
following  theorem (for whose proof we demand to
\cite{Reed-Simon-79}):
\begin{theorem}
\end{theorem}
ON THE UNIQUENESS OF REPRESENTATIONS OF WEYL RELATION:

\begin{hypothesis}
\end{hypothesis}
\begin{center}
 $ \{ \hat{V}_{1}(t) \}_{t \in {\mathbb{R}} } ,  \{ \hat{V}_{2}(s) \}_{s \in {\mathbb{R}} } $ one parameter strongly-continuous unitary group
 on a separable Hilbert space $ {\mathcal{H}} $ satisfying the \emph{Weyl relation}
\end{center}
\begin{thesis}
\end{thesis}
There are closed linear subspaces $ {\mathcal{H}}_{l} $ such that:
\begin{itemize}
    \item
\begin{equation*}
   {\mathcal{H}} \; = \; \oplus_{l=1}^{N} {\mathcal{H}}_{l} \; \;
   N \in {\mathbb{N}}_{+} \cup \{ \infty \}
\end{equation*}
    \item
\begin{equation*}
    \hat{U}(t) :  {\mathcal{H}}_{l} \mapsto {\mathcal{H}}_{l} \, , \,
    \hat{V}(s) : {\mathcal{H}}_{l} \mapsto {\mathcal{H}}_{l} \; \;
    \forall s,t \in {\mathbb{R}}
\end{equation*}
    \item
\begin{equation*}
    \forall l , \exists \hat{T}_{l} : {\mathcal{H}}_{l} \mapsto L^{2}( {\mathbb{R}}  ,
{\mathbb{C}}) \; unitary \; \; : \; \hat{T}_{l} \hat{U}(t)
\hat{T}_{l}^{- 1}  =  \exp ( i t \hat{P}_{q} ) \; and \;
\hat{T}_{l} \hat{V}(s) \hat{T}_{l}^{- 1}  =  \exp ( i s
\hat{Q}_{q} )
\end{equation*}
\end{itemize}

It is anyway possible to insist on working with the
\emph{Canonical Commutation Relation} provided one adds further
hypotheses to the Conjecture \ref{con:naive version of Von Neumann
Uniqueness Theorem} under which it becomes a theorem.

The first step in this direction is rather trivial, consisting
simply in getting rid of the reducibility of representations:

given a dense linear subspace $ {\mathcal{D}} $ of $ {\mathcal{H}}
$:
\begin{definition}
\end{definition}
$ \hat{\tilde{Q}} $ AND $  \hat{\tilde{P}} $ ARE A SELF-ADJOINT
IRREDUCIBLE REPRESENTATION OF THE CANONICAL COMMUTATION RELATION
OVER $ {\mathcal{D} } $

\begin{itemize}
    \item $ \hat{\tilde{Q}} $ and $  \hat{\tilde{P}} $ are
    self-adjoint
    \item $ \hat{\tilde{Q}} $ and $  \hat{\tilde{P}} $ are a
    representation of the \emph{Canonical Commutation Relation}
    over $ {\mathcal{D}} $
    \item
\begin{equation*}
    \nexists  {\mathcal{I}} \text{ dense linear subspace of $ {\mathcal{D}} $
    } \; : \; ( \exp ( i s \hat{\tilde{Q}} ) {\mathcal{I}} \subseteq
    {\mathcal{I}} \, \forall s \in {\mathbb{R}} )   \; or \; ( \exp ( i t \hat{\tilde{P}} ) {\mathcal{I}} \subseteq
    {\mathcal{I}} \, \forall t \in {\mathbb{R}} )
\end{equation*}
\end{itemize}

Among the many possibilities one is the following
\cite{Summers-01}:

\begin{theorem} \label{th:Von Neumann Uniqueness Theorem in streghtened Dixmier's form}
\end{theorem}
VON NEUMANN UNIQUENESS THEOREM (IN WEAKENED DIXMIER'S FORM)

\begin{hypothesis}
\end{hypothesis}
\begin{center}
  $ \hat{\tilde{Q}} $ and $  \hat{\tilde{P}} $ are a self-adjoint irreducible
  representation of the \emph{Canonical Commutation Relation} over a
  dense linear subspace  $ {\mathcal{D}} $ of an
  Hilbert space $ {\mathcal{H}} $ such that $ \hat{\tilde{Q}} $ and $  \hat{\tilde{P}} $
  are closed  and the restriction of $ \hat{\tilde{Q}}^{2} +
  \hat{\tilde{P}}^{2} $ to $ {\mathcal{D}} $ is
  essentially self-adjoint
\end{center}
\begin{thesis}
\end{thesis}
\begin{equation*}
    \exists \hat{U} :  {\mathcal{H}} \mapsto  L^{2}( {\mathbb{R}} ,
{\mathbb{C}}) \text{ unitary } \; : \; \hat{\tilde{Q}} =
\hat{U}^{- 1} \hat{Q}_{q} \hat{U} \; and \;  \hat{\tilde{P}} =
\hat{U}^{- 1} \hat{P}_{q} \hat{U}
\end{equation*}

\smallskip

The \emph{Schr\"{o}dinger representation} of the \emph{Canonical
Commutation Relation} is also  called the \emph{position
representation}.

Let us now introduce the following operators:
\begin{equation}
    ( \hat{Q}_{p} \psi ) ( p) \; := \; + i \frac{d \psi (p)}{ d p}
\end{equation}
\begin{equation}
     ( \hat{P}_{p} \psi ) ( p) \; := \; p  \psi(p)
\end{equation}
defined as the closures of their restriction to the initial domain
$ {\mathcal{S}} ( {\mathbb{R}} , {\mathbb{C}}) $.

It may be easily verified that $ \hat{Q}_{p} $ and $ \hat{P}_{p} $
 are an irreducible self-adjoint representation of the
\emph{Canonical Commutation Relation}, called the \emph{momentum
representation}, over a dense linear subspace of $ L^{2}(
{\mathbb{R}} , {\mathbb{C}}) $ over which $ \hat{Q}_{p}^{2} +
\hat{P}_{p}^{2} $ is essentially self-adjoint.

Applying  Theorem\ref{th:Von Neumann Uniqueness Theorem in
streghtened Dixmier's form} it follows that:
\begin{corollary} \label{cor:on the unitarily equivalence of the position and momentum representation}
\end{corollary}
ON THE UNITARILY  EQUIVALENCE OF THE POSITION AND MOMENTUM
REPRESENTATIONS
\begin{equation*}
    \exists \hat{U} :  {\mathcal{H}} \mapsto  L^{2}( {\mathbb{R}} ,
{\mathbb{C}}) \text{ unitary } \; : \; \hat{Q}_{p} = \hat{U}^{- 1}
\hat{Q}_{q} \hat{U} \; and \;  \hat{P}_{p} = \hat{U}^{- 1}
\hat{P}_{q} \hat{U}
\end{equation*}

Indeed the unitary of Corollary \ref{cor:on the unitarily
equivalence of the position and momentum representation} is
nothing but the usual (complex) Fourier transform.

\newpage
\section{Position and momentum representations of the Hyperbolic Canonical Commutation Relation}
Given the \emph{ Hyperbolic Canonical Commutation Relation}:
\begin{equation}
    [ \hat{q} , \hat{p} ] \; = \; j \hat{I}
\end{equation}
let us consider its \emph{position representation} in $ L^{2} (
{\mathbb{R}} , {\mathbb{G}})$ :
\begin{equation}
    ( \hat{q}_{q} \psi ) ( q) \; := \; q \psi (q)
\end{equation}
\begin{equation}
     ( \hat{p}_{q} \psi ) ( q) \; := \; - j  \frac{d \psi (q)}{ d q}
\end{equation}
and its \emph{momentum representation} in $ L^{2} ( {\mathbb{R}} ,
{\mathbb{G}})$ :
\begin{equation}
    ( \hat{q}_{p} \psi ) ( p) \; := \; + j \frac{d \psi (p)}{ d p}
\end{equation}
\begin{equation}
     ( \hat{p}_{p} \psi ) ( p) \; := \; p  \psi(p)
\end{equation}
where all the operators are defined as the closures of their
restriction on the initial domain $ {\mathcal{S}} ( {\mathbb{R}} ,
{\mathbb{G}}) $.

We will prove the following:
\begin{theorem} \label{th:on the unitarily inequivalence of the hyperbolic position and momentum representation}
\end{theorem}
ON THE UNITARILY  INEQUIVALENCE OF THE HYPERBOLIC POSITION AND
MOMENTUM REPRESENTATIONS

$ \nexists \hat{U} : L^{2} ( {\mathbb{R}} , {\mathbb{G}})  \mapsto
L^{2} ( {\mathbb{R}} , {\mathbb{G}}) \text{ unitary } \, : $
\begin{equation*}
  \hat{q}_{p} \; = \; \hat{U}^{- 1} \hat{q}_{q} \hat{U}
\end{equation*}
\begin{equation*}
  \hat{p}_{p} \; = \; \hat{U}^{-1} \hat{p}_{q} \hat{U}
\end{equation*}
\begin{proof}
Owing to theorem \ref{th:no hyperbolic Plancherel theorem} we know
that the required $ \hat{U} $ is not the Fourier transform as
instead occurs in Quantum Mechanics.

 Let us now follow for a moment the non-rigorous Dirac bra-ket
formalism.

Starting with:
\begin{equation}
    < q | \hat{q} | \alpha > \; = \; q < q | \alpha >
\end{equation}
\begin{equation}
    < q | \hat{p}  | \alpha >  \; = \; - j \frac{d }{dq} < q |
    \alpha >
\end{equation}
one has in particular that:
\begin{equation}
 < q | \hat{p}  | p >  \; = \; - j \frac{d }{dq} < q | p >
\end{equation}
and hence:
\begin{equation}
  [ j \frac{d }{dq} + p ] < q | p > \; = \; 0
\end{equation}
from which it follows that:
\begin{equation} \label{eq:braket between position autobra and momentum autoket}
  < q | p > \; = \; c \exp ( - j p q ) \; \; c \in {\mathbb{G}}
\end{equation}
But then one has that:
\begin{multline} \label{eq:exchanging representation}
    < p |  \hat{q} | \alpha > \; = \; \int_{- \infty}^{+ \infty}
    dq < p | q > < q | \hat{q} | \alpha > \; = \bar{c} \int_{- \infty}^{+ \infty}
    dq \exp ( j p q ) < q | \hat{q} | \alpha > \; = \\
     \bar{c} \; \int_{- \infty}^{+ \infty}
    dq \exp (  j p q ) q < q | \alpha > \; = \; \frac{\bar{c}}{ j}
    \frac{d}{dp} \int_{- \infty}^{+ \infty}
    dq \exp (  j p q ) < q | \alpha > \; = \\
      j  \frac{d}{dp} \int_{- \infty}^{+ \infty}
    dq < p | q > < q | \alpha >  \; = \;  \; + j \frac{d}{dp} < p |
    \alpha >  \; = \; ( \hat{q}_{p} \psi_{|\alpha >} ) (p)
\end{multline}
where we have used eq.\ref{eq:braket between position autobra and
momentum autoket} and  the completeness condition for position
autokets:
\begin{equation}
    \int_{- \infty}^{+ \infty}
    dq  | q > < q | \; = \; \hat{I}
\end{equation}

Eq.\ref{eq:exchanging representation} implies that:
\begin{equation}
  \hat{q}_{p} [ f ]\; = \; ({\mathcal{F}} \hat{q}_{q}
  {\mathcal{F}}^{- 1} ) [f] \; \; \forall f \in  N [ {\mathcal{F}} ]
\end{equation}
\begin{equation}
   \hat{p}_{p} [ f ]\; = \; ({\mathcal{F}} \hat{p}_{q}
  {\mathcal{F}}^{- 1} ) [f] \; \; \forall f \in  N [ {\mathcal{F}} ]
\end{equation}
where:
\begin{equation}
    N [ {\mathcal{F}} ] \; := \; \{ f \in D ( {\mathcal{F}} )   \, : \, {\mathcal{F}} [f] \in  {\mathcal{S}} ( {\mathbb{R}} ,
    {\mathbb{G}}) \}
\end{equation}
So, assuming ad absurdum the existence of a unitary $ \hat{U} :
L^{2} ( {\mathbb{R}} , {\mathbb{G}})  \mapsto L^{2} ( {\mathbb{R}}
, {\mathbb{G}}) $ :
\begin{equation}
  \hat{q}_{p} \; = \; \hat{U}^{-1} \hat{q}_{q} \hat{U}
\end{equation}
\begin{equation}
  \hat{p}_{p} \; = \; \hat{U}^{-1} \hat{p}_{q} \hat{U}
\end{equation}
, one should have that $ \hat{U}|_{N[ {\mathcal{F}} ]}  \; = \;
{\mathcal{F}}^{- 1} $.
\end{proof}

\smallskip

The same formulation of a conjecture claiming the existence an an
analogous of Theorem\ref{th:Von Neumann Uniqueness Theorem in
streghtened Dixmier's form} for operators on an hyperbolic Hilbert
space would be an highly not trivial task owing to the
peculiarities of self-adjoint operators on such a space discussed
in the appendix \ref{sec:Self-adjoint operators on an hyperbolic
Hilbert space}.

Theorem\ref{th:on the unitarily inequivalence of the hyperbolic
position and momentum representation}, anyway, automatically
implies that such a conjecture would be false, i.e. that Von
Neumann Uniqueness Theorem doesn't hold in Hyperbolic Quantum
Mechanics.

In fact, if an hyperbolic quantum mechanical analogous of
theorem\ref{th:Von Neumann Uniqueness Theorem in streghtened
Dixmier's form} existed, it would imply the violation of theorem
\ref{th:on the unitarily inequivalence of the hyperbolic position
and momentum representation}.
\newpage
\appendix
\section{Hyperbolic functions at rapid decrease and hyperbolic tempered distributions}
\begin{definition}
\end{definition}
HYPERBOLIC FUNCTIONS OF RAPID DECREASE:
\begin{equation*}
    {\mathcal{S}} ( {\mathbb{R}} ,  {\mathbb{G}} ) \; := \; \{ f : {\mathbb{R}} \rightarrow  {\mathbb{G}} \text{ infinitely differentiable } :
     \| f \|_{n, m}  :=
      \sup_{x \in {\mathbb{R}}} \|
     x^{n} \frac{d^{m}}{ d x^{m}} f(x) \| < \infty
      \; \forall n , m \in {\mathbb{N}}_{+} \}
\end{equation*}
Let us endow $ {\mathcal{S}} ( {\mathbb{R}} ,  {\mathbb{G}} ) $
with the natural topology induced by the seminorms $ \| \cdot
\|_{n , m} $.
\begin{definition}
\end{definition}
SPACE OF HYPERBOLIC TEMPERED DISTRIBUTIONS:
\begin{equation*}
      {\mathcal{S}}' ( {\mathbb{R}} ,  {\mathbb{G}} ) \; := \;
      \text{ topological-dual } [ {\mathcal{S}} ( {\mathbb{R}} ,  {\mathbb{G}}
      ) ]
\end{equation*}

In particular let us introduce the following:
\begin{definition}
\end{definition}
HYPERBOLIC DIRAC DELTA:
\begin{equation*}
    \delta \in {\mathcal{S}}' ( {\mathbb{R}} ,  {\mathbb{G}} ) \;
    :   \delta [ f ]  := f(0)
\end{equation*}

Given an hyperbolic tempered distribution $ \lambda \in
{\mathcal{S}}' ( {\mathbb{R}} , {\mathbb{G}} ) $, a family of
functions $ f_{\alpha} : {\mathbb{R}} \rightarrow {\mathbb{G}}  $
for every $\alpha \in I:= [a,b] $ with $ a,b \in [ 0 , + \infty ]
$, a measure $ \mu $ on $ ( {\mathbb{R}} , {\mathcal{B}} (
{\mathbb{R}} ))$ and a number $ \bar{\alpha} \in I $:
\begin{definition}
\end{definition}
 $ f_{\alpha} $ IS A LIMIT-REPRESENTION OF $ \lambda $ WITH RESPECT TO $
 \mu $ FOR $ \alpha \rightarrow  \bar{\alpha} \; \;  ( REP_{\mu}-\lim_{ \alpha \rightarrow \bar{\alpha}}
  f_{\alpha} = \lambda ) $
 \begin{equation}
    \lim_{ \alpha \rightarrow \bar{\alpha} } \int d \mu (x)
    f_{\alpha} (x) \phi (x) \; = \;  \lambda [ \phi ]  \; \;
    \forall  \phi \in {\mathcal{S}} ( {\mathbb{R}} ,  {\mathbb{G}} )
\end{equation}
Let us now consider the family of functions:
\begin{equation}
    f_{\alpha} (x) := \int_{- \alpha}^{\alpha} dp \exp ( j p x )
    \; = \; \frac{2 \sinh ( \alpha x ) }{x}
\end{equation}
One has that:
\begin{theorem}
\end{theorem}
\begin{equation*}
     REP_{\mu_{Lebesgue}}-\lim_{ \alpha \rightarrow + \infty
 } f_{\alpha} \neq \delta
\end{equation*}
\begin{proof}
Let us consider the test function $ \phi (x) := \exp ( - x^{2} )
$.

Since:
\begin{equation}
    \int_{- \alpha}^{\alpha} dp \exp ( j p x - x^{2})
\end{equation}
doesn't converge to $ \phi (0) = 1 $ as $ \alpha \rightarrow
\infty $ the thesis follows
\end{proof}

\smallskip

One has that:
\begin{theorem} \label{th:embedding theorem}
\end{theorem}
EMBEDDING THEOREM
\begin{equation*}
     {\mathcal{S}} ( {\mathbb{R}} ,  {\mathbb{G}}
      ) \; \subset \;  L^{2} ( {\mathbb{R}} ,  {\mathbb{G}}
      ) \; \subset \; {\mathcal{S}}' ( {\mathbb{R}} ,  {\mathbb{G}}
      )
\end{equation*}
\begin{proof}
Every function $ f \in {\mathcal{S}} ( {\mathbb{R}} , {\mathbb{G}}
) $ can be identified with the functional $ f[ \cdot ] \in
{\mathcal{S}}' ( {\mathbb{R}} , {\mathbb{G}} ) $ defined as:
\begin{equation}
    f [g] \: := \; \int_{- \infty}^{+ \infty} dx g(x) f(x)
\end{equation}
From the other side one has that:
\begin{equation}
   ( \| f \|_{n,m} < + \infty \, \forall n,m \in {\mathbb{N}}_{+} ) \; \Rightarrow \;  \int_{- \infty}^{+ \infty}
    dx \| f(x) \|^{2} \in ( - \infty , + \infty )
\end{equation}
and hence $ f \in L^{2} ( {\mathbb{R}} , {\mathbb{G}} ) $
\end{proof}
\newpage
\section{The hyperbolic Fourier transform}
Let us introduce the following:
\begin{definition}
\end{definition}
HYPERBOLIC FOURIER TRANSFORM:

the functional $ {\mathcal{F}} : D( {\mathcal{F}} ) \mapsto
MAPS({\mathbb{R}}, {\mathbb{G}} ) $:
\begin{equation*}
    D( {\mathcal{F}} ) \; := \; \{ f \in {\mathcal{S}} ( {\mathbb{R}} ,  {\mathbb{G}} )
    \,
: \, \exists \text{ finite }  \int_{-\infty}^{+ \infty} dx \exp (
- j p x)
    f(x) \}
\end{equation*}
\begin{equation*}
   ( {\mathcal{F}}[f])( p ) \; := \;  \int_{-\infty}^{+ \infty} dx \exp ( - j p x) f(x)
\end{equation*}
Let us observe that:
\begin{proposition} \label{prop:tempered distributions are not invariant under Fourier transform}
\end{proposition}
\begin{equation*}
  {\mathcal{F}} [  D( {\mathcal{F}} ) ] \;
  \nsubseteq \; L^{2} ( {\mathbb{R}} ,  {\mathbb{G}})
\end{equation*}
\begin{proof}
Given the function $ f(x) := \frac{1}{\sqrt{2 \pi}} \exp ( -
\frac{x^{2}}{2}) \in {\mathcal{S}} ( {\mathbb{R}} ,  {\mathbb{G}})
$ one has that:
\begin{equation}
    {\mathcal{F}} [ f] (p) \; = \; \exp ( \frac{p^{2} }{2} ) \;
    \notin \; L^{2} ( {\mathbb{R}} ,  {\mathbb{G}})
\end{equation}
\end{proof}

Proposition \ref{prop:tempered distributions are not invariant
under Fourier transform} implies that:
\begin{theorem} \label{th:no hyperbolic Plancherel theorem}
\end{theorem}
NO HYPERBOLIC PLANCHEREL THEOREM:

\emph{$ {\mathcal{F}} $ doesn't extend to a unitary $ \hat{U}
 : L^{2} ( {\mathbb{R}} ,  {\mathbb{G}}) \mapsto L^{2} (
{\mathbb{R}} , {\mathbb{G}}) $}
\newpage
\section{Self-adjoint operators on an hyperbolic Hilbert space} \label{sec:Self-adjoint operators on an hyperbolic Hilbert space}

Let $ ( X , \| \cdot \|_{X} ) $ and $ ( Y  , \| \cdot \|_{Y} ) $
be hyperbolic Banach spaces. Given a bounded linear operator $ T :
X \mapsto Y $ let us introduce the following:
\begin{definition}
\end{definition}
HYPERBOLIC BANACH SPACE ADJOINT OF T:

the operator $ T' : Y^{\star} \mapsto X^{\star}$ :
\begin{equation}
   (T' l) (x) \; := \; l (T x) \; \; l \in Y^{\star} , x \in X
\end{equation}

Let us now consider an hyperbolic Hilbert space $ ( {\mathcal{H}}
, ( \cdot , \cdot ) , \| \cdot \| )$. The hyperbolic Banach space
adjoint of a bounded linear operator $ T : {\mathcal{H}} \mapsto
{\mathcal{H}} $ is then an operator $ T' : {\mathcal{H}}^{\star}
\mapsto {\mathcal{H}}^{\star} $.

Let us now consider the map $ C : {\mathcal{H}} \mapsto
{\mathcal{H}}^{\star} $ which assigns to each $ y \in
{\mathcal{H}} $ the linear functional $ ( y , \cdot ) \in
{\mathcal{H}}^{\star}$.

The key difference with respect to the analogous situation on an
Hilbert space \cite{Reed-Simon-80} is that in our case there is no
analogue of Riesz Lemma and hence nobody assures us that C is
surjective.

We are thus led to the following:
\begin{definition}
\end{definition}
HYPERBOLIC HILBERT SPACE ADJOINT OF T:

the operator $ T^{\dag} : D(T^{\dag}) \mapsto {\mathcal{H}} $:
\begin{equation}
    T^{\dag} \; := \; C^{- 1} T' C
\end{equation}
where $  D(T^{\dag}) $ is the linear subspace of $  {\mathcal{H}}
$ where $ C^{- 1} T' C $ is well-defined.

The notion of hyperbolic Hilbert space adjoint can be extended to
unbounded operators in the following way:

given an unbounded linear operator T defined on a dense subspace
D(T) of an hyperbolic Hilbert space $ {\mathcal{H}} $:
\begin{definition}
\end{definition}
HYPERBOLIC HILBERT SPACE ADJOINT OF T:

the operator $ T^{\dag} : D(T^{\dag}) \mapsto {\mathcal{H}} $:
\begin{equation*}
    D( T^{\dag} ) \; := \; \{ \phi \in  {\mathcal{H}} \; : \; (
    \exists ! \eta_{\phi} \in  {\mathcal{H}} : ( T \psi , \phi ) =
    ( \psi , \eta_{\phi} ) \: \forall \psi \in D(T) ) \}
\end{equation*}
\begin{equation*}
   T^{\dag} \phi \; := \; \eta_{\phi}
\end{equation*}
\begin{definition}
\end{definition}
T IS SELF-ADJOINT:
\begin{equation*}
    D(T) =  D( T^{\dag} ) \; and \; T = T^{\dag}
\end{equation*}

\smallskip

A standard theorem of Functional Analysis  asserts that the
spectrum of a self-adjoint operator on an Hilbert space is a
subset of the real line \cite{Reed-Simon-80}.

This is no more  true as to self-adjoint operators on a hyperbolic
Hilbert space as we will show in the simplest case $ {\mathcal{H}}
= {\mathbb{G}}^{2}$.

Let us introduce at this purpose the  following useful bijection $
T :  M_{n } ( {\mathbb{G}} )  \mapsto  M_{n } ( {\mathbb{C}} ) $:
\begin{equation}
    T ( \{ x_{i,j}+ j y_{i,j} \}_{i,j=1}^{n} ) \; := \; \{ x_{i,j}+ i y_{i,j} \}_{i,j=1}^{n}
\end{equation}
 Given:
\begin{equation}
    A \; := \; \left(%
\begin{array}{cc}
  x_{11} + j y_{11} &  x_{12} + j y_{12} \\
  x_{21} + j y _{21}& x_{22} + j y_{22} \\
\end{array}%
\right) \in  M_{2 } ( {\mathbb{G}} )
\end{equation}
one has that:
\begin{equation}
    A^{\dag} \; = \; \bar{A}^{t} \; = \;  \left(%
\begin{array}{cc}
  x_{11} - j y_{11} &  x_{21} -j y_{21} \\
  x_{12} - j y _{12}& x_{22} -j  y_{22} \\
\end{array}%
\right)
\end{equation}
Let us introduce the set of self-adjoint matrices:
\begin{equation}
    SA_{2} ( {\mathbb{G}} ) \; := \; \{ A \in M_{2 } ( {\mathbb{G}}
    ) \, : \, A^{\dag} = A \}
\end{equation}
Clearly one has that:
\begin{equation}
    A \in  SA_{2} ( {\mathbb{G}} )  \; \Leftrightarrow \; y_{11}=
    0 \, and \,  y_{22} = 0 \, and \, x_{12} =  x_{21} \, and \,  y_{12} = - y_{21}
\end{equation}
so that the generic matrix $  A \in  SA_{2} ( {\mathbb{G}} )$ is
of the form:
\begin{equation}
    A \; = \; \left(%
\begin{array}{cc}
  x_{11} & x_{12} + j y_{12} \\
  x_{12} - j y_{12}  & x_{22} \\
\end{array}%
\right)
\end{equation}
Let us compare the eigenvalue equation of A and T(A). For the
latter the equation:
\begin{equation}
    det( T(A) - \lambda I) \; = \; 0
\end{equation}
has solution:
\begin{equation}
    \lambda \; = \; \frac{ x_{11}+x_{22} \pm \sqrt{( x_{11}-x_{22})^{2}   + 4 x_{12}^{2} + 4 y_{12}^{2} }}{2}
\end{equation}
and since the discriminant $ \Delta := ( x_{11}-x_{22})^{2}   + 4
x_{12}^{2} + 4 y_{12}^{2} \; \geq 0 $:
\begin{itemize}
    \item if $ \Delta  := ( x_{11}-x_{22})^{2}   + 4
x_{12}^{2} + 4 y_{12}^{2} > 0 $ then  T(A) has 2 real eigenvalues:
\begin{equation}
    \lambda_{1} = \frac{ x_{11}+x_{22} + \sqrt{\Delta}}{2} \in {\mathbb{R}}
\end{equation}
\begin{equation}
    \lambda_{2} = \frac{ x_{11}+x_{22} - \sqrt{\Delta }}{2} \in {\mathbb{R}}
\end{equation}
    \item if  $ \Delta  := ( x_{11}-x_{22})^{2}   + 4
x_{12}^{2} + 4 y_{12}^{2} = 0 $ then  T(A) has 1 real eigenvalue:
\begin{equation}
    \lambda = \frac{ x_{11}+x_{22}}{2} \in {\mathbb{R}}
\end{equation}
\end{itemize}
As to A, instead, the equation:
\begin{equation}
    det( A - \lambda I) \; = \; 0
\end{equation}
has solution:
\begin{equation}
    \lambda \; = \; \frac{ x_{11}+x_{22} \pm \sqrt{( x_{11}-x_{22})^{2}   + 4 x_{12}^{2} - 4 y_{12}^{2} }}{2}
\end{equation}
If follows that:
\begin{itemize}
    \item if $ \Delta := ( x_{11}-x_{22})^{2}   + 4
x_{12}^{2} - 4 y_{12}^{2} \; > 0 $ then A has 4 eigenvalues of
which only two are reals:
\begin{equation}
    \lambda_{1} = \frac{ x_{11}+x_{22} + \sqrt{\Delta }}{2} \in {\mathbb{R}}
\end{equation}
\begin{equation}
    \lambda_{2} = \frac{ x_{11}+x_{22} - \sqrt{\Delta }}{2} \in {\mathbb{R}}
\end{equation}
\begin{equation}
    \lambda_{3} = \frac{ x_{11}+x_{22} + j \sqrt{\Delta }}{2} \notin {\mathbb{R}}
\end{equation}
\begin{equation}
    \lambda_{4} = \frac{ x_{11}+x_{22} - j \sqrt{\Delta }}{2} \notin {\mathbb{R}}
\end{equation}
 \item if  $ \Delta := ( x_{11}-x_{22})^{2}  + 4
x_{12}^{2}  - 4 y_{12}^{2} \; = \; 0 $ then A has 1 real
eigenvalue:
\begin{equation}
    \lambda = \frac{ x_{11}+x_{22}}{2} \in {\mathbb{R}}
\end{equation}
 \item if  $ \Delta := ( x_{11}-x_{22})^{2}  + 4
x_{12}^{2}  - 4 y_{12}^{2} \; < \; 0 $ then A has no eigenvalues
\end{itemize}

In particular we have shown that a matrix $ A \in SA_{2}
({\mathbb{G}}) $ cannot be always diagonalized, a fact that by
itself  proves that the Spectral Theorem doesn't hold for
self-adjoint operators on an hyperbolic Hilbert space.

This fact implies  that given a self-adjoint operator A on an
hyperbolic Hilbert space:
\begin{itemize}
    \item if A is bounded, the exponential of A can be defined by power-series:
\begin{equation}
    \exp ( j t A) \; := \; \sum_{n=0}^{\infty} \frac{(jt)^{n} A^{n}}{n !}
\end{equation}
    \item if A is unbounded, not only the exponential of A cannot
    be defined by power series (as occurs also for an unbounded
    operator on a (complex) Hilbert space), but one cannot  use the
    functional calculus form of the Spectral theorem; it follows
    that no definition of $ \exp ( j t A) $ can be given in this
    way.
\end{itemize}
As a consequence it follows that no analogue exists on an
Hyperbolic Hilbert Space of the Stone Theorem that on a (complex)
Hilbert space states the existence of a bijection between
self-adjoint operators and strongly-continuous unitary groups
associating to each self-adjoint operator A the
strongly-continuous unitary group $ \{  \exp ( i t A) \}_{t \in
{\mathbb{R}}} $.

\end{document}